# SmartFPS: Neural Network based Wireless-inertial fusion positioning system

Luchi Hua, Jun Yang

*Abstract*—**The current fusion positioning systems are mainly based on filtering algorithms, such as Kalman filtering or particle filtering. However, the system complexity of practical application scenarios is often very high, such as noise modeling in pedestrian inertial navigation systems, or environmental noise modeling in fingerprint matching and localization algorithms. To solve this problem, this paper proposes a fusion positioning system based on deep learning and proposes a transfer learning strategy for improving the performance of neural network models for samples with different distributions. The results show that in the whole floor scenario, the average positioning accuracy of the fusion network is 0.506m. The experiment results of transfer learning show that the estimation accuracy of the inertial navigation positioning step size and rotation angle of different pedestrians can be improved by 53.3% on average, the Bluetooth positioning accuracy of different devices can be improved by 33.4%, and the fusion can be improved by 31.6%.**

## I. INTRODUCTION

Location navigation service is one of the indispensable technologies for modern society and scientific development. However, the current mature GPS positioning is usually unable to locate effectively indoors due to irregular attenuation caused by the occlusion of GPS signals by clouds, building walls, and ceilings. Because of this, the new technology of indoor positioning system was proposed. At present, mainstream indoor positioning technologies include Wi-Fi [1-3], Bluetooth [4-6], UWB [7,8], Lidar [9,10], machine vision [11,12], inertial navigation [13-15], Visible light [16-18] and so on. Each positioning technique has its advantages as well as its limitations. For example, inertial navigation positioning is prone to accumulative errors due to system noise and drift [19, 20]; positioning signals such as Wi-Fi and Bluetooth fluctuate and signal attenuation is difficult to model, so traditional methods such as trilateral positioning [21, 22] are directly used for positioning accuracy. Not high; UWB has a high cost [23], so it is difficult to promote it in the consumer field; Lidar has a high cost and has certain requirements for wall reflection coefficient [24], so it has poor performance in low-light environments; and jitter [25] have certain requirements; visible light is easily occluded [26] so the positioning is discontinuous. In general, high-precision, and high-stable positioning performance cannot be obtained based on a single positioning system. On the other hand, high-cost positioning systems cannot be used for civilian use. Especially in the pedestrian positioning scenario, factors such as portability and cost need to be considered. Therefore, most positioning solutions obtain multiple sensor data from the user's mobile terminal to coordinate positioning services, such as the gyroscope and accelerometer of the inertial unit in the smartphone. Wi-Fi and Bluetooth modules, etc. Due to the limitations of various positioning technologies, fusion positioning systems have been widely studied in recent years, such as Wi-Fi, Bluetooth, Lidar, and inertial navigation fusion [27-32]. Simply put, data fusion is the process of combining data from multiple sensors and related information to achieve more specific inferences than can be achieved with a single sensor. However, for pedestrian localization, the current mainstream filter fusion based systems are very cumbersome to implement and have poor performance. Because the fusion system not only needs to consider the model parameters and noise characteristics of each sub-positioning system, especially the pedestrian motion characteristics, but also needs to consider the fusion algorithm parameters and the applicability of various algorithms, so the implementation is more complicated and suitable for different scenarios. There is a big difference in performance. To further improve the positioning accuracy and stability of pedestrians, it is urgent to explore a more advanced fusion system.

In the actual pedestrian location scenario, the following key issues need to be considered: 1) Multi-signal: the receiver may receive zero, one or more beacon signals at the same time; 2) Occlusion: due to different signal penetration attenuation characteristics, such as super The penetration of high-frequency signals is poor, and when there are walls or pedestrians passing in front of the receiver, one or more direct signals are blocked; 3) Diversity of tracking states: several variables may be tracked at the same time in positioning, Such as position, velocity, angle, etc.; 4) Nonlinear system model: The system model is generally nonlinear, such as the relationship between position, time and acceleration under non-uniform motion; 5) Nonlinear measurement model: such as the wireless signal measured by the receiver Intensity, it needs to be transformed into the system space, and according to the signal attenuation model, it is transformed into a nonlinear process; 6) Non-Gaussian noise: The system often has non-Gaussian noise, such as the rate drift of inertial sensors; 7) Continuity of positioning: The position and velocity of the target (i.e., the state space) can be estimated continuously. The current fusion positioning system for pedestrian positioning is mainly based on filtering algorithms, such as Kalman filtering or particle filtering. Compared with the single-system indoor positioning system, the fusion algorithm has higher stability. However, in many practical application scenarios, the complexity of the fusion system or a single system is already very high, such as pedestrian positioning for the modeling of inertial sensor signal estimation step angle, or the dependence of Bluetooth signal modeling on the environment. At this time, it is difficult for the traditional fusion system to model a single positioning model or its noise model, and the increase of the system complexity will make the parameter error transmitted to the final state estimation error larger. Therefore, many studies apply machine learning techniques to the modeling of a single positioning system, but the fusion positioning is still based on filtering algorithms. Taking the indoor pedestrian localization scene based on Bluetooth as an example, this topic proposes a fusion localization system based on deep learning and proposes a transfer learning strategy for improving the performance of the neural network model for samples with different distributions. Collect tagged data and improve the positioning accuracy of the system.

## II. RELATED WORKS

### A. Pedestrian Inertial Navigation

Pedestrian inertial navigation mainly includes strapdown inertial navigation system [76] and step size direction estimation system (SHS) [77-79].

A typical strapdown inertial navigation system mainly includes the following steps:
1) Calculate the gyroscope bias, accelerometer bias and positioning state error including position error, velocity error and rotation angle error by extended Kalman filter.
2) Correct the current inertial device output through the gyroscope deviation and accelerometer deviation.
3) The current rotation angle is obtained by the integration of the gyroscope output and the accumulation of the rotation angle error.
4) Convert the acceleration data from the inertial system coordinate system to the global coordinate system through the current rotation angle.
5) The current speed is obtained by the integration of the accelerometer output after the coordinate system conversion and the accumulation of the speed error.
6) The displacement value is obtained through the current speed, and the final position information is obtained by combining the position error.

If there is an error in the angle estimation, the final position result will be affected by exponential amplification through multi-layer transfer. The effect of this error is especially severe for low-cost MEMS sensors in smartphones. The use of extended Kalman filtering can play a certain role in suppressing this error.

To deal with drift, it is necessary to close the integrating loop periodically by imposing external constraints on the system. Currently the most widely used constraint method is the zero velocity update (ZUPT) [81]. ZUPT is based on the sensor being at rest and can be applied during the stance phase, provided the sensor is attached to the foot. ZUPT is easily incorporated into the INS structure by representing ZUPT as a pseudo-measurement of zero velocity. The application of ZUPT means that the open loop integration only occurs during the swing phase of the foot connected to the sensor. For such short durations, the accumulation of drift is limited, so longer tracking durations are feasible. However, for reliable output, ZUPT must only be applied when the foot (and therefore the sensor) is completely stationary. Problems can arise when the sensor is mounted higher than the ball of the foot. The peeling motion associated with the transition from standing to swinging means that the heel rises soon after the foot down event, so the sensor in the midfoot will begin to experience acceleration as the foot lifts. These small accelerations occur before the strict end of the stance phase, so it is necessary to account for these errors by applying a non-zero covariance next to the ZUPT pseudo-measurement.

To estimate pedestrian step size and direction more accurately, research [82] proposed an inertial navigation system based on recurrent neural network [83]. This system is different from the traditional SHS system, the estimation of step size and direction is estimated by the model trained on the data. Since the neural network is very good at estimating nonlinear systems, the step size and direction errors estimated by the model trained on the data are smaller. The network used in this system is a long short-term memory neural network (LSTM) [84], which is a special neural network structure and is mainly designed for time series data. The input of the system is the accelerometer gyroscope time sequence under a certain time window, the network structure is a double-layer bidirectional recurrent neural network, and the output is the step size and deflection angle of the pedestrian movement in the time window. On the premise that the initial position and direction of the pedestrian are known, the system estimates the position and direction of each step through the estimated step size and deflection angle. In this study, it is verified by experiments that the neural network-based method has a significant improvement in positioning accuracy compared with the traditional PDR method. The method based on neural network not only improves the positioning accuracy, but also simplifies various complex processes such as data processing and noise analysis.

Based on IONet, research [85] made further improvements. The preprocessing of inertial data is analyzed in detail in this study, and the collected data is based on the random walking posture of pedestrians in the larger environment of the entire floor. In terms of data preprocessing, this study introduces the spatiotemporal calibration process of the data in detail. Because the collected data comes from two different mobile phones, the mobile phone for position acquisition is fixed in front of the body, while the mobile phone for inertial data collection moves freely with the arm, so although the relative distance and angle between the two mobile phones will be constrained Within a certain range, but other motion data of the mobile phone can be used to convert the inertial data to the global coordinate system. In this study, three network structures were investigated, including ResNet network, LSTM network, and TCN network of convolutional neural network. Another difference from the study is that the output of the system proposed in this study is a velocity vector. Through experiments to verify the results, when the research algorithm is applied to the data set of this study, the positioning accuracy drops significantly, and the positioning accuracy of the algorithm proposed in this study is much higher than the method of the study [82].

### B. Indoor Positioning based on Wireless Signals

Wireless signal-based positioning systems generally build positioning models based on features such as received signal strength, signal arrival angle, and signal time of flight (ToF), but most existing wireless-based positioning solutions mainly rely on RSS-based features, which have the advantage of Simple to implement. RSS-based wireless positioning algorithms are mainly divided into three types, namely proximity [65], trilateration and fingerprinting.

Proximity positioning mainly judges whether the receiving device is close to the wireless beacon according to the RSS value, to directly use the position of the beacon to estimate the position of the receiving device. The proximity positioning system has high stability. Although the wireless RSS will fluctuate, when the device is very close to the wireless beacon, the RSS value after a certain smoothing process is within a range. Proximity positioning is mainly used in scenarios that do not require high positioning accuracy. To achieve continuous high-precision positioning, many beacons need to be deployed. Proximity positioning needs to consider the setting of the threshold and the distance between the wireless beacons.

Wireless RSS-based trilateration is a positioning algorithm that estimates the location of a device based on the RSS values of three or more wireless beacons. Due to the fluctuation of wireless RSS, the circles of three distances in trilateral positioning do not intersect at the same point, but the overlapping part of the three circles is an area. To estimate the position of the receiver, this position can be calculated by nonlinear optimization methods, such as least squares.

Obviously, for trilateral positioning, the positioning accuracy depends on at least three stable direct wireless signals. In general positioning scenarios, due to the consideration of deployment costs, the deployment distance of wireless beacons is large, plus the interference or occlusion of external noise, it is difficult to continuously collect three stable direct wireless signals, so trilateral positioning has certain limitations in the wireless positioning system.

To solve the problem of wireless positioning under low signal-to-noise ratio, fingerprint positioning technology is more and more widely used. Fingerprint positioning technology mainly collects RSSI measurement values from different wireless beacons at different locations in the scene in an offline manner, and the combination of these measurement values becomes the fingerprint information of the location. After the system is deployed, the RSSI combination of online measurements (obtained in real time) is compared with offline measurements to estimate user location. Fingerprint localization algorithms usually need to survey the environment to obtain fingerprints or characteristics of the environment in which the localization system is used, so it is also called scene analysis localization technology.

### C. Fusion Algorithms based on filters

The filtering algorithm is one of the most widely used fusion algorithms in the current mainstream indoor fusion positioning systems. At present, the filtering algorithms applied in the positioning fusion algorithm mainly include: 1) discrete Bayesian filter; 2) Kalman filter; 3) particle filter. The Kalman filter also includes extended Kalman filter, unscented Kalman filter and so on.

The Kalman filter is theoretically the best estimate for unimodal linear systems with Gaussian noise, but not for non-Gaussian nonlinear systems. Therefore, in view of the limitations of the Kalman filter algorithm, various improved Kalman filter algorithms such as extended Kalman filter and unscented Kalman filter were proposed in subsequent research.

Since the extended Kalman filter is a first-order estimation for nonlinear systems, ideal localization performance cannot be obtained for strongly nonlinear systems. In response to this limitation, there are also Kalman filters for second order and third-order estimates. In addition, extended Kalman filtering cannot guarantee the convergence of the algorithm. If the initial state quantity error is large, or the process model is incorrect, the algorithm will diverge. In addition, there is an error between the covariance matrix of the extended Kalman filter algorithm and the actual covariance matrix, so continuous localization performance cannot be guaranteed.

UKF can handle nonlinear, continuous, multivariate problems. Sigma points can also estimate a certain degree of non-Gaussian noise, but cannot be accurately estimated for complex non-Gaussian distribution problems.

Particle filtering can handle nonlinearity and non-Gaussian noise, but at different degrees of nonlinearity and non-Gaussian noise, the number of particles, the generation strategy and the resampling strategy have a high impact on the accuracy, and more particles will reduce the calculation speed. In addition, the performance of particle filtering in dealing with high-dimensional systems is not good, because high-dimensional systems can easily lead to excessive differences in the weight distribution of particles, resulting in the loss of particle diversity.

In addition to the above filtering algorithms, many studies have also proposed other well-known algorithms including Integrated Kalman Filter (EnKF) [91], Adaptive Kalman Filter (AKF) [92], Switched Kalman Filter (SKF) [93].

## III. SYSTEM ARCHITECTURE

### A. System Overview

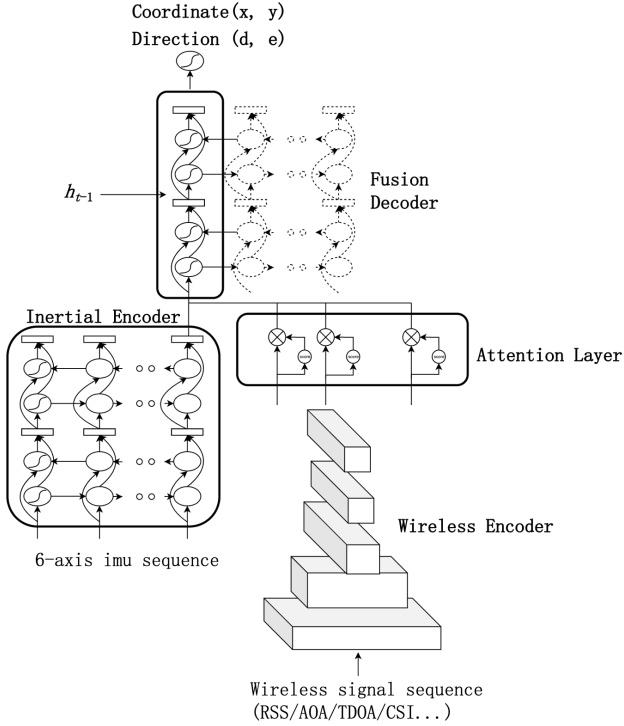

Fig. 1 System Overview

The wireless inertial navigation fusion positioning system based on neural network uses the accelerometer gyroscope signal sequence and the wireless signal sequence of the inertial navigation system as input, and finally estimates the current pedestrian position and direction. The structure of the fusion positioning system is shown in Fig. 1. The system mainly includes four parts: feature extraction module of inertial navigation data based on long short-term memory network, wireless positioning module based on convolutional neural network, attention layer and fusion positioning module based on long short-term memory network. Among them, the inertial navigation data feature extraction module and the wireless positioning module need to be pre-trained. The output of the inertial navigation data feature extraction module is the latent space tensor of step size and steering angle, and the output of the wireless positioning module is also the extracted wireless signal feature information rather than the final position information. The outputs of the two modules pass through the attention layer. Input to the fusion localization module. The fusion positioning module using the hidden state of the previous moment as the initial state finally estimates the position and direction information.

Compared with the filtering fusion positioning system, the fusion positioning system based on the whole network structure has the following advantages:

1) The fusion positioning system with full network structure can realize end-to-end computing output, which is easier to train and deploy than the positioning system with network plus filtering algorithm. Each positioning network of the network plus filtering algorithm needs to output a noise matrix during training, and the noise matrix is difficult to represent by individual parameters for unstable systems. Therefore, such training and fusion methods are complicated, and the implementation steps are cumbersome.

2) Compared with Kalman filtering, the fusion positioning system with full network structure needs to assume that all noises are white noise, and no need to pre-determine the noise coefficient. It can approximate any white noise or colored noise through data training and will eventually learn into the system. stable noise, and fault tolerance for unstable noise.

3) The fusion positioning system with full network structure is faster than particle filtering, and the time interval of model output results is generally in milliseconds, while the time cost of particle filtering according to the number of particle swarms is much higher than that of the network model, and the particle Filtering will cause problems such as loss of diversity over time, so parameters such as particle resampling need to be adjusted.

### B. Inertial Encoder

The pre-training network of the inertial navigation coding network in the fusion positioning network is an inertial navigation feature extraction network based on long short-term memory network. Its input is the 6-axis accelerometer gyroscope signal in the inertial unit of the smart device, and the signal data set at a moment can be expressed as ($acc_x$, $acc_y$, $acc_z$, $gyro_x$, $gyro_y$, $gyro_z$). The long short-term memory network generally processes a time series of data, in this research scenario, the accelerometer gyroscope data under a time window. The length of the window can be selected according to different application scenarios. For example, in this study, a time window of 1 s is used for the frequency of pedestrian steps, which represents the time required for about two steps.

The reason why the data of the accelerometer and the gyroscope are selected in the input data of the inertial unit of the smart device is mainly because the fusion of the accelerometer and the gyroscope can make the accumulated results of the accelerometer and the gyroscope smoother. The main characteristic of accelerometers is that they are sensitive to vibrations yet remain stable over long periods of time. While the gyroscope is not sensitive to vibration, it will drift seriously when used for a long time. When estimating the deflection angle, since the gyroscope generally drifts seriously in a few minutes or encounters a large inertial force, and the accelerometer has higher stability, for the low-cost inertial sensor of the mobile phone device, the accelerometer will the time frequency that needs to be calibrated is generally around a day, so the accelerometer can be used for gyroscope drift calibration. Sometimes, when correcting the yaw angle, the magnetometer and accelerometer gyroscope are often used for fusion. However, because the magnetometer on the mobile phone is easily affected by geomagnetism and changes in different geographical locations, it is not as stable as the accelerometer unless special methods are used to suppress the disturbance. To prevent the disturbance of the magnetometer from affecting the result of the overall fusion calculation, generally, the magnetometer may not be used to fuse with other sensors in the pedestrian inertial navigation positioning.

The output of the inertial navigation feature extraction network is the step size and the deflection angle. Although for a set of continuous time sequences, the position information can be directly calculated by the accelerator gyroscope in the traditional PDR algorithm, but the step size and deflection angle are used as the output here. The inertial navigation feature extraction network is completely decoupled from the map, which is very necessary for machine learning algorithms. In the study of RoNIN [85], the author used the velocity vector as the output. In fact, there is a serious logical problem, that is, when the pedestrian keeps walking in a straight line at a constant speed, the velocity vector should be possible in all directions, so that the trained network is easy to overfit. The author randomly rotates the inertial data for data enhancement when training the network, which is helpful for decoupling, but this problem is not clearly pointed out in the text.

The inertial navigation feature extraction network is a two-layer long short-term memory neural network, in which each time result output by the first layer is output to the second layer network, and the output of the second layer network is the hidden state quantity at the last moment. A fully connected layer finally outputs two values of step size and deflection angle. The long short-term memory network is a bidirectional type, that is, the entire inertial data time series can be considered globally, so that the data characteristics at a certain moment also affect the data characteristics at the previous moment.

The objective function of the inertial positioning network is as follows:

$$\min_\theta L(\theta) = \frac{1}{n}\sum_{i=1}^{n} L(x_i, y_i; \theta) = \frac{1}{n}\sum_{i=1}^{n} \left( \left\| \tilde{l} - l \right\|_2^2 + \lambda \left\| d\tilde{\varphi} - d\varphi \right\|_2^2 \right) \quad (1)$$

where $n$ represents the number of batch data, $\theta$ represents the network parameters, $l$ represents the step size, $d\varphi$ represents the deflection angle, and $\lambda$ is the weight learned by the deflection angle network. The training target of the entire network includes two parts: step size and deflection angle. Due to the different physical meanings of the two parts, the range of values is also different. For example, the maximum value of the step size is related to the pedestrian step size and the selected time window size. will remain in the range of [0,1], and the value of the deflection angle in this study is in the range of (-180°, 180°], so two target value ranges should be considered for the setting of the weight, and the Consider training speed.

*C. Wireless Signal Encoder*

The wireless positioning module is also called the wireless encoder, and its main function is to map the input wireless signal features to the latent space. Different from previous studies, this study uses dynamically collected wireless time-series signals as input to predict 2D position results. The input feature is a feature matrix composed of the signal sequence of the wireless beacons deployed in the entire environment within a time window. The matrix length is the window size, the height is the number of wireless beacons, and the thickness is 1, that is, the received signal strength value. The output 2D position is the time window midpoint value. The size of the time window depends on two factors, generalization performance and training accuracy. For example, the use of an excessively long-time window should be avoided, because the dynamic acquisition of signal data using a long-time window will produce timing dependencies, thereby reducing the generalization performance of the wireless location network. Second, due to the high noise of the wireless signal received by the smartphone, if the time window is too short, the network learns the noise signal instead of the signal itself, which will lead to overfitting of the network. Therefore, the size of the time window is a game of two aspects, and the optimal solution can be found by changing the positioning accuracy of the validation set.

The wireless coding network is built based on the convolutional neural network. Although the wireless signal is collected dynamically, and the dynamic sequence generally pays attention to its timing relationship, the purpose of this network is to use the network to learn the signal itself in the high-noise signal sequence rather than the timing relationship, which can be regarded as a noise reduction process for high-noise signals. Therefore, the selection of convolutional neural network is based on global feature considerations. It is also found in the experiment that the performance difference between the use of the cyclic neural network structure and the convolutional neural network is small, and even the test performance is not as good as the convolutional neural network. The convolutional neural network of the wireless coding network mainly includes three convolutional layers, a maximum pooling layer, a global average pooling layer and two fully connected layers.

The fully connected layer is divided into two types according to the activation function as a linear function or a nonlinear function. Generally, for regression algorithm models, the last layer is often a linear fully connected layer, that is, a simple addition of the input of the pre-layer. The front fully connected layer is generally a structure driven by a nonlinear activation function, and its main function is to characterize the characteristics of the input. As the system complexity increases, the number of layers and hidden units of the nonlinear fully connected layer tends to increase.

The size of the convolution kernel of each layer of the convolutional neural network gradually decreases, that is, from global attention to local attention, and the step size also selects different values for tensors with different length and width. The specific network hyperparameters will be detailed in the appendix List. The loss function of the wireless location network is as follows:

$$\min_\theta L(\theta) = \frac{1}{n}\sum_{i=1}^{n} L(x_i, y_i; \theta) = \frac{1}{n}\sum_{i=1}^{n} \sqrt{(\tilde{x} - x)^2 + (\tilde{y} - y)^2} \quad (2)$$

where $x, y$ are the two-dimensional position coordinates. There is no need to add a regularization term to the loss function. The main consideration is that the uncertainty of the position, that is, $x$ and $y$, will change with the specific environment of different locations in the actual scene and the deployment of wireless beacons, so the parameters of the regularization term are more difficult to determine.

It is worth noting that this network uses the selu function as the activation function for all convolutional neural networks. The selu function is expressed as follows:

$$\begin{cases} a = \lambda z & , z > 0 \\ a = \lambda \alpha \left( e^z - 1 \right), z \le 0 \end{cases} \quad (3)$$

Compared with the relu activation function, selu has the advantage of not having a dead zone. Compared with sigmoid, it also has the advantage that the gradient is not easy to disappear. Compared with the elu function, it has the advantage of no parameter selection. The most important role of the selu function is that it can automatically normalize the sample distribution to 0 mean and unit variance, thereby speeding up the network convergence. The function is like the batch normalization layer, but there is no need to increase the network depth. We also found that the selu function can improve the Training speed is also important for online training and domain adaptation algorithms. Although the number of iterations for the final convergence is basically close, the validation set of the wireless location network has converged in the initial stage, so using the selu function is very useful to speed up the training of this network, especially when retraining.

### D. Asymmetric Attention Layer

Asymmetric attention layers are mainly based on the concept of local attention in deep learning [95]. Local attention is a type of attention mechanism. Attention mechanism is a concept proposed in the field of natural language processing, which is mainly used to implement a contextual logic between the encoding network and the decoding network. The attention mechanism generally includes three steps: calculating the calibration weight through the hidden state, calculating the softmax weight, and calculating the overall context weight. The concept of local attention is relative to the global attention. The global attention implements the attention mechanism for all units of the input sequence, while the local attention mechanism implements the attention mechanism for some units. The use of both mechanisms depends on whether the part of the latent vector sequence that needs attention is all or part of it.

In the wireless inertial navigation fusion positioning system, the attention mechanism is used for the output of the wireless coding network, but the attention mechanism is not used for the inertial navigation coding network. This mechanism is called asymmetric attention mechanism, and its essence is local attention. Mechanism that allows the output of the inertial navigation coding network to pass through. For the wireless inertial navigation fusion positioning system, the hidden state output by the wireless coding network and the hidden state output by the inertial navigation coding network itself have different physical meanings. Since the output of the inertial navigation coding network is the hidden state quantity of the last step, it does not have the concept of context itself. If the output of the inertial navigation coding network adopts the attention mechanism, it will break the integrity of the part of the hidden state, and only input the part of the hidden state sequence into the decoding network. As for the wireless coding network, due to the different positions of each wireless beacon, when pedestrians are in different positions, the signal strength and stability of each wireless beacon are also different. At the same time, due to factors such as occlusion by walls or multipath effects, it cannot be determined whether to use the signal of the beacon as an input simply by the strength of the signal. Therefore, the attention mechanism can enable the localization network to learn the local characteristics of the environment and realize the regional selection of appropriate wireless signal characteristics, especially when the number of wireless beacons is large, so it is convenient to extend the positioning system in practical applications. The calculation process of the asymmetric attention layer is as follows:

$$u_t = \tanh(Wh_t + b) \tag{4}$$

$$\alpha_t = \text{softmax}(u_t) \tag{5}$$

$$s = \sum_{t=1}^{M} \alpha_t h_t \tag{6}$$

where $W$ refers to the weight of the fully connected layer in the attention layer, $b$ is the bias of the fully connected layer, and $h_t$ is the amount of hidden state input to the attention layer. Compared with the global attention mechanism, the asymmetric attention mechanism greatly reduces the number of network parameters, which is of great significance for suppressing model overfitting and improving model training and computing speed.

### E. Fusion System Decoder

The fusion decoding network is a single-step and one-way long-term and short-term memory network structure based on two layers. The hidden state output of the attention layer is used as the input of the network, and the hidden state and unit state of the network are initialized to the hidden state of the previous moment. The structure of the fusion decoding network is shown in Figure 3-10. The hidden state quantities of the input attention layer include inertial navigation features and attention-selected wireless positioning signal features, which can be regarded as observations at the current moment, so the role of the recurrent neural network here is like a data-driven filtering process. However, the recurrent neural network can selectively fuse the two observational features and predict the state of the current moment. The advantage of the recurrent neural network is that the noise feature of the observation at the current moment is used as the hidden state input together with the observation, and the state quantity and the noise feature of the previous moment are also used as the hidden state to initialize the network unit. State quantities, observations, and system processes and noise are coupled through the network, thereby eliminating the tedious process of determining system noise. Especially for the wireless location system based on fingerprint location, the noise at different locations is more difficult to quantify. The hidden state quantity output by the fusion decoding network will be further connected to two fully connected layers, and the output of the last fully connected layer is the two-dimensional position coordinate vector (x, y) and direction vector (rx, ry) of the pedestrian. The target of the model is set as the root mean square error of the two-dimensional coordinate vector and the direction vector, with a scale factor of $\kappa$.

The final objective function of the fusion decoding network is as follows:

$$\min_{\theta} L(\theta) = \sum L(x, y; \theta) = \sum \left\| \tilde{d} - d \right\|_2^2 + \kappa \left\| \tilde{r} - r \right\|_2^2 \tag{7}$$

where $\theta$ is the network parameter, $d$ represents the two-dimensional position vector (x, y), $r$ represents the direction vector (rx, ry).

### F. Training method based on Multi-task Learning

Multi-task Learning [96] is a branch of deep learning that aims to improve learning efficiency by exploiting the similarity between different tasks to solve multiple different tasks simultaneously. Formally, if there are n tasks (traditional deep learning methods aim to solve only 1 task using 1 specific model), where these n tasks or subsets of them are related to each other but not identical, multi-task learning is achieved by Using the knowledge contained in all n tasks will help improve the learning of a particular model. Multi-task learning gives us the possibility to further improve performance by forcing the model to learn a more general representation as it learns (updates its weights). Intuitively, biologically humans learn in the same way. It is possible to learn better if you learn multiple related tasks rather than focusing on one specific task for a long time. Multi-task learning is used today in many fields, such as object detection and facial recognition, self-driving cars (which can detect pedestrians, stop signs, and other obstacles simultaneously), multi-domain collaborative filtering for web applications, stock prediction, language modeling and other NLP applications.

In the training of the fusion positioning network, we use different goals for each sub-network for training, which itself is a learning process of multiple tasks, but this process is a single-threaded step-by-step process. Here we propose a fusion network training method of multi-task learning, which is to introduce the remaining part of the pre-training model of the sub-network into the fusion positioning network at the same time, and at the same time retain the training targets of each sub-network. The multi-task learning network is with the following goals:

$$\min_\theta L(\theta) = \min_\theta \left( L_{fusion}(\theta) + \lambda_1 L_{ins}(\theta) + \lambda_2 L_{wireless}(\theta) \right) \quad (8)$$

Among them, represents the weight of the inertial navigation network target, and represents the weight of the wireless positioning network target. Although multi-task learning is generally used to improve learning efficiency, it can play a role in optimizing the integrity of the output features of the intermediate layer in this study, and the fixed data feature positions can not only retain the physical meaning of the output of each sub-network, but also play a role in the later paper. Domain adaptation systems based on generative adversarial networks play a very important role.

## IV. SMARTFPS TRANSFER LEARNING

### A. Attack Factors

To analyze the influence of different pedestrians on the characteristics of the inertial navigation data, we use the same equipment, and select two different testers except the training set collectors to collect the test data. The three persons walked back and forth along the same straight-line path, and the mobile phone holding pose was similar to that of the training set. Fig. 2 shows the time series of the accelerometer z-axis acceleration data of the acquisition equipment of the training set collector (height 1.76m) and one of the testers (height 1.63m). The left side is the data collected by the training set, and the right side is the data collected by the tester. From the characteristics of time series data, it can be found that the peak-to-peak ratio of the z-axis acceleration of the training set collectors is greater than that of the testers, that is, the mobile phone swings up and down with greater force. In addition, it can be found that there are frequent sub-peaks near the peak in the tester's data, which is also caused by the tester's exercise habits. Fig. 3 shows the time sequence of the rotation rate of the mobile phone gyroscope around the z-axis when the trainer and the tester walk in a straight line. It can be observed from the figure that the peak-to-peak value of the data of the training personnel is smaller than that of the test personnel. During the experimental collection process, it is also found that the left and right swing amplitude of the test personnel is higher when holding the mobile phone. Although in a general inertial navigation positioning system, the estimation of the step size and the deflection angle can be regarded as being mainly obtained by integrating the inertial data in the horizontal direction, but the disturbance caused by different motion characteristics also affects the components of the inertial data in the horizontal direction. will have an impact. Therefore, most of the pedestrian inertial navigation positioning research needs to add the analysis of pedestrian motion characteristics to the system. However, it is difficult to accurately model the diversity of the characteristics and changes of different pedestrian movements, so it is difficult to accurately track the position coordinates of pedestrians by relying solely on the pedestrian inertial navigation positioning system. For the inertial navigation wireless fusion positioning system in this study, it is necessary to consider how to strengthen the generalization performance of the original network under this characteristic attack.

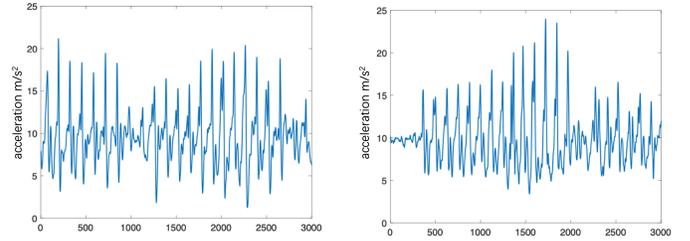

Fig. 2 Signal sequence of accelerator along z-axis. (a) person 1; (b) person 2.

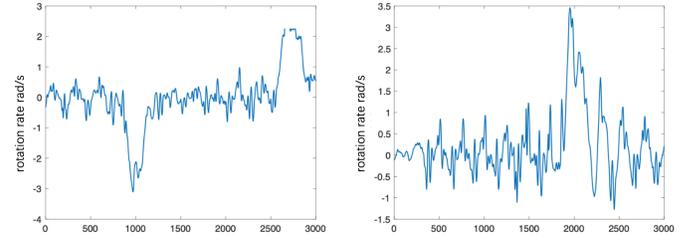

Fig. 3. Allan variance of gyroscope along z-axis. (a) person 1; (b) person 2.

In addition to the influence of pedestrian motion characteristics on data characteristics, different devices also cause data changes. Because the method of initial calibration of the inertial system can eliminate problems such as bias and drift, we only consider the noise characteristics of the inertial device itself of different devices. In order to verify the difference in noise characteristics of inertial systems of different devices, this study uses two devices (Xiaomi Mi6 and ZTE A2017) to collect inertial data in a horizontal stationary state. For the signal timing of the dynamic state, please refer to Section 5.1.2, but since it is difficult for the two devices to be relatively stationary when they are in dynamic motion, we do not perform noise analysis on them. To analyze the noise, Allan's variance method was used in this study. The Allan variance curves of the accelerometer and gyroscope sequences of the two devices in the static state are shown in Fig. 4 and Fig. 5. From Fig. 4, it can be observed that the main noise characteristics of the accelerometer of Mi6 and A2017 within 1 second are Gaussian white noise, but the white noise coefficients are quite different, and the A2017 is close to 1s. There are small magnitude colored noise components. Therefore, the noise characteristics of accelerometers are quite different between inertial devices of different equipment. The noise is diverse within 1s, and there is obvious random walk noise above 1s.

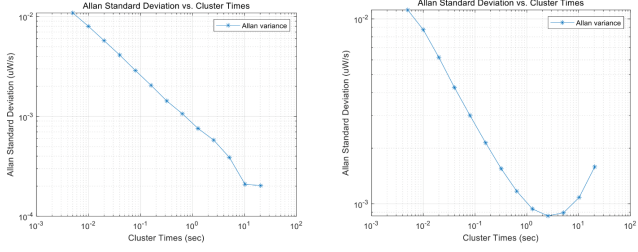

Fig. 4 Allan variance of accelerator along x-axis. (a) Mi 6; (b) ZTE A2017.

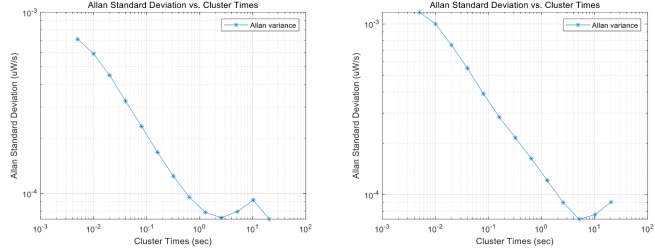

Fig. 5 Allan variance of gyroscope along x-axis. (a) Mi 6; (b) ZTE A2017.

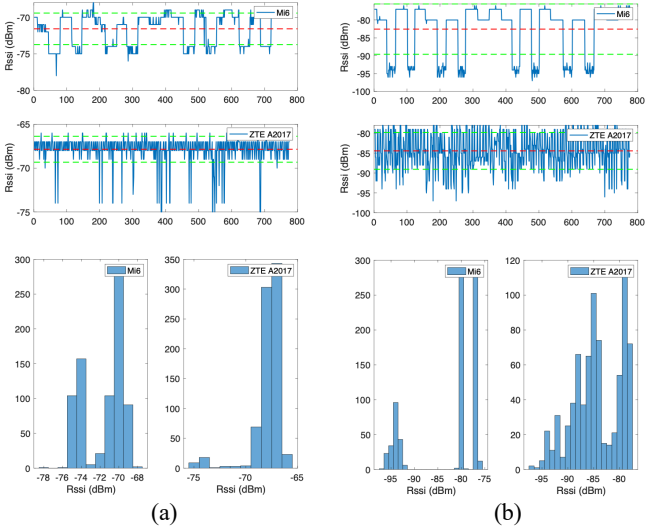

Fig. 6 The received signal strength sequence of a Bluetooth beacon under two different distances. (a) 0.6m; (b) 1.5m.

There are also differences in the wireless signal receivers of different mobile phone devices. To verify this problem, in this study, two types of mobile phones were placed horizontally on the table and kept the same distance from the Bluetooth beacon to collect signal strength data. Fig. 6 shows the signal strength data when the distance from the Bluetooth beacon is 0.6m and 1.5m. Fig. 6(a) shows the Bluetooth signal strength data of the two devices at 0.6m. The above figure of Fig. 6(a) is the signal timing diagram, in which the middle-dotted line represents the mean, and the upper and lower dotted lines represent the mean plus or minus the standard deviation. The bottom figure of Fig. 6(a) shows the statistical histograms of different signal strength values. It can be observed that the signal data noise of the two devices is quite different, the A2017 signal has a higher frequency of change, and the Mi6 has a lower signal change frequency. In addition, although the distance between the two devices and the Bluetooth beacon is the same, there is a certain difference in the received signal strength values. The average value of the Mi6 is about -72dBm, and the A2017 is about -68dBm. From the histogram, there are two obvious peaks in the Bluetooth signal data histogram of Mi6, and the amplitude values are obviously discrete. The A2017 noise frequency is higher, but the amplitude change is smaller. There are many outliers in A2017, and the problem can be clearly seen from the timing diagram. Fig. 6(b) shows the signal strength data at a distance of 1.5m. It can be seen from the timing diagram that the signal of A2017 still has the characteristics of high frequency noise. The histogram shows that the two peaks of the Mi6's Bluetooth signal are farther apart, and the A2017 also begins to have two peaks.

To sum up, there are irregular differences in the amplitude and noise characteristics of the wireless signal strength data sampled by the wireless receivers of different devices. Therefore, it is necessary to consider how to enhance the generalization performance of the wireless inertial navigation fusion positioning network under this difference.

## B. Transfer Learning based on GAN

The domain adaptation algorithm of the wireless positioning network is mainly aimed at the training of the first 3-layer CNN network layer in the wireless positioning network, and the wireless positioning encoder in the fusion network model of multi-task learning. The main function of the wireless encoder is to map the features of the wireless signal data sequence, so the main purpose of the local adaptation algorithm is to train this layer of network and substitute the trained weights into the fusion positioning network to improve the target. Domain fusion localization performance. The domain adaptation network structure of the wireless positioning network is mainly based on the structure of the generative adversarial network. The role of the generator is to generate the wireless sequence of another domain according to the wireless positioning features of one domain.

We also add a Cycle-consistent goal to the network, so the generator network will include a source domain generator and a target domain generator. These two parts share one unique encoder. According to the logic of cycleGAN, the source domain generator network needs to restore the sequence of false target domain signals generated by the target domain generator (the false target domain signals are also derived from the source domain data) sequence to the source domain data. Similarly, the target domain generator network needs to restore the sequence of fake target domain signals generated by the source domain generator (fake source domain signals are generated from target domain data) to target domain data. Such a mechanism makes the features not unduly influenced by the generator network when the matching degree of the source and target data labels is low.

In addition, a new loss function, the identity loss, is often introduced when applying cycleGAN. This mechanism means that data in the same domain should maintain the same data through the same domain generator network. This mechanism is one step shorter than the path length of cycle consistency loss (this step includes generating fake data through another domain generator network and passing. The encoder extracts features in two sub-steps).

Inspired by the reconstruction network, we introduce the reconstruction network in the domain adaptation network. Its structure is a CNN structure, and its role is to restore the original signal using the encoder output features. Unlike the consistency loss mechanism, the reconstruction network is unique. Although the recovered signal will not be exactly the same as the original, it preserves its signal integrity as much as possible from the common characteristics of the source and target domains. This part is also to protect the encoder from being unduly influenced by the generative network. Since the adversarial generative network needs to train the generator and the discriminator separately, we also divide the training of the inertial positioning network domain adaptation network into two steps, which correspond to two goals: 1) include training the generator 2) the discriminator target. The network forward propagation route of the two targets is shown in Fig. 7.

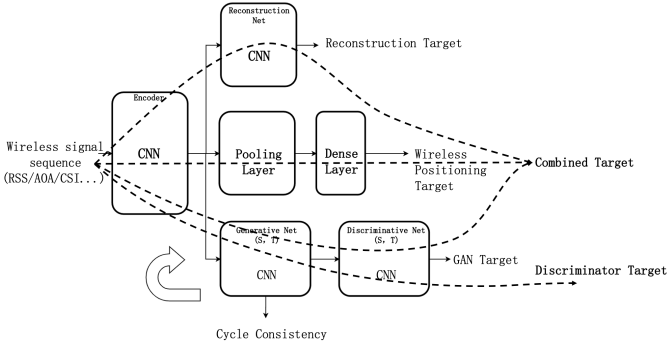

Fig. 7 Transfer learning structure of wireless positioning encoder in SmartFPS.

The wireless positioning loss, reconstruction loss, consistency loss and cycle consistency loss are the training of the generator and the encoder, and the weights of the discriminator network are frozen during the training process, so they can be combined to form a combined target through weight distribution. The function is (take the source domain as an example):

$$L_{c,S} = L_{gan} + \lambda_1 L_{cycle} + \lambda_2 L_{identity} + \lambda_3 L_{pred} + \lambda_4 L_{recon} \quad (9)$$

The training procedure of wireless positioning encoder transfer learning is similar to train GAN. It is important to train only the generator or the discriminator alone while freeze the weights of the other network. However, the other goals can be trained together. The procedure is listed in Table 1.

The transfer learning structure of inertial encoder is similar to the wireless positioning encoder, only the network class should be replaced by LSTM structure.

After training the wireless positioning encoder and inertial encoder seperately, the weights of each encoder can be used in the fusion network so that the general transfer learning target can be accomplished.

**Table 1 Wireless Positioning Encoder Domain Adaptation Algorithm**

1) Get weights of the wireless positioning network of the fusion positioning network trained by multi-task learning.
2) Initialize the wireless positioning network weights in the domain adaptation network, and freeze the pooling layer and dense layer weights.
3) Network training:
for t=1→Number of iterations:
    Sampling N labeled data from the source domain;
    Sampling N unlabeled data from the target domain;
    The generator network generates target domain fake data from source domain data;
    The generator network generates target domain fake data from target domain data;
    Freeze the source domain discriminator network weights and update the combined network weights according to the source domain combined target:

$$\nabla_{\theta_F, \theta_{G_S}, \theta_C} \sum_{i=1..N} L_{gan} \left( \bar{D}_S \left( G_S \left( F \left( x_{T,i}, \theta_F \right), \theta_{G_S} \right), \theta_{D_S} \right), y_{real} \right) +$$

$$\lambda_1 \begin{pmatrix} L_{cycle} \left( \bar{G}_T \left( F \left( G_S \left( F \left( x_{T,i}, \theta_F \right), \theta_{G_S} \right), \theta_F \right), \theta_{G_T} \right), x_{T,i} \right) + \\ L_{cycle} \left( G_S \left( F \left( \bar{G}_T \left( F \left( x_{S,i}, \theta_F \right), \theta_{G_T} \right), \theta_F \right), \theta_{G_S} \right), x_{S,i} \right) \end{pmatrix} +$$

$$\lambda_2 L_{identity} \left( G_S \left( F \left( x_{S,i}, \theta_F \right), \theta_{G_S} \right), x_{S,i} \right) + \lambda_3 L_{pred} \left( \bar{R} \left( F \left( x_{S,i}, \theta_F \right), \theta_R \right), y_{S,i} \right)$$

$$+ \lambda_4 \left( L_{recon} \left( C \left( F \left( x_{S,i}, \theta_F \right), \theta_C \right), x_{S,i} \right) + L_{recon} \left( C \left( F \left( x_{T,i}, \theta_F \right), \theta_C \right), x_{T,i} \right) \right)$$

    Freeze the INS coding network, the generator network, and update the discriminator network weights according to the source domain discriminator target:

$$\nabla_{\theta_{D_S}} \sum_{i=1..N} \begin{pmatrix} L_{gan} \left( D_S \left( x_{S,i}, \theta_{D_S} \right), y_{real} \right) + \\ L_{gan} \left( D_S \left( \bar{G}_S \left( \bar{F} \left( x_{T,i}, \theta_F \right), \theta_{G_S} \right), \theta_{D_S} \right), y_{fake} \right) \end{pmatrix}$$

    Freeze the target domain discriminator network weights, and combine the targets according to the target domain: Update the combined network weights;

$$\nabla_{\theta_F, \theta_{G_T}, \theta_C} \sum_{i=1..N} L_{gan} \left( \bar{D}_T \left( G_T \left( F \left( x_{S,i}, \theta_F \right), \theta_{G_T} \right), \theta_{D_T} \right), y_{real} \right) +$$

$$\lambda_1 \begin{pmatrix} L_{cycle} \left( \bar{G}_S \left( F \left( G_T \left( F \left( x_{S,i}, \theta_F \right), \theta_{G_T} \right), \theta_S \right), \theta_{G_S} \right), x_{S,i} \right) + \\ L_{cycle} \left( G_T \left( F \left( \bar{G}_S \left( F \left( x_{T,i}, \theta_F \right), \theta_{G_S} \right), \theta_F \right), \theta_{G_T} \right), x_{T,i} \right) \end{pmatrix} +$$

$$\lambda_2 L_{identity} \left( G_T \left( F \left( x_{T,i}, \theta_F \right), \theta_{G_T} \right), x_{T,i} \right) + \lambda_3 L_{pred} \left( \bar{R} \left( F \left( x_{S,i}, \theta_F \right), \theta_R \right), y_{S,i} \right)$$

$$+ \lambda_4 \left( L_{recon} \left( C \left( F \left( x_{T,i}, \theta_F \right), \theta_C \right), x_{T,i} \right) + L_{recon} \left( C \left( F \left( x_{S,i}, \theta_F \right), \theta_C \right), x_{S,i} \right) \right)$$

    Freeze the INS coding network, the generator network, and update the discriminator network weights according to the target domain discriminator target:

$$\nabla_{\theta_{D_T}} \sum_{i=1..N} \begin{pmatrix} L_{gan} \left( D_T \left( x_{T,i}, \theta_{D_T} \right), y_{real} \right) + \\ L_{gan} \left( D_T \left( \bar{G}_T \left( \bar{F} \left( x_{S,i}, \theta_F \right), \theta_{G_T} \right), \theta_{D_T} \right), y_{fake} \right) \end{pmatrix}$$

end for

## V. Experiment

### A. Simulation

In order to compare the performance of data fusion and filter fusion algorithms based on recurrent neural network, this paper realizes two types of fusion systems based on extended Kalman filter and particle filter through the output of two pre-trained models of wireless and inertial navigation. The input of the wireless positioning network takes the Bluetooth signal propagation model as an example, and the received signal strength value is generated mainly through the position label of the inertial navigation data set of IONet. Five Bluetooth beacons are evenly distributed in the simulation environment.

Among them, for the extended Kalman filter, the variance matrix of the observed amount of the Bluetooth estimated position is determined by the pre-training network using the variance value of the error between the prediction result of the training set and the actual value. Let the state quantity be , because there is the following relationship between the two-dimensional position vector and the step deflection angle:

$$\begin{cases} x_{k+1} = x_k + l_k \cos \varphi_k \\ y_{k+1} = y_k + l_k \sin \varphi_k \end{cases} \quad (10)$$

Therefore, the process equation for the state quantity is derived from the above formula:

$$\begin{cases} dx_{k+1} = dx_k + dl_k \cos \varphi_k - l_k d\varphi_k \sin \varphi_k + dw_x \\ dy_{k+1} = dy_k + dl_k \sin \varphi_k + l_k d\varphi_k \cos \varphi_k + dw_y \\ dl_{k+1} = dl_k + dw_l \\ d\varphi_{k+1} = d\varphi_k + dw_\varphi \end{cases} \quad (11)$$

Therefore, the process matrix is:

$$\Phi_k = \begin{bmatrix} 1 & 0 & \cos \varphi_k & -l_k \sin \varphi_k \\ 0 & 1 & \sin \varphi_k & l_k \cos \varphi_k \\ 0 & 0 & 1 & 0 \\ 0 & 0 & 0 & 1 \end{bmatrix} \quad (12)$$

The observation equation is:

$$Z = [dx_k, dy_k]^T = [x_{ble,k} - x_{pred,k}, y_{ble,k} - y_{pred,k}] \quad (13)$$

So the observation matrix is:

$$H_k = \begin{bmatrix} 1 & 0 & 0 & 0 \\ 0 & 1 & 0 & 0 \end{bmatrix} \quad (14)$$

The process noise matrix is:

$$Q_k = \begin{bmatrix} \delta_x^2 & 0 & 0 & 0 \\ 0 & \delta_y^2 & 0 & 0 \\ 0 & 0 & \delta_l^2 & 0 \\ 0 & 0 & 0 & \delta_\varphi^2 \end{bmatrix} \quad (15)$$

The observation noise matrix is:

$$R_k = \begin{bmatrix} R_x & 0 & 0 & 0 \\ 0 & R_y & 0 & 0 \end{bmatrix} \quad (16)$$

The implementation process of the particle filter system is mainly based on the process equation, which will not be repeated here. The particle weight update is based on how close the particle position is to the output value of the Bluetooth network. We set the number of particles to 700 as suggested in [60]. It is worth noting that how to set the position noise in the process noise has a great influence on the positioning accuracy of the filtering system. However, since the position and step deflection angle information output by the positioning network is used in this simulation as the state quantity for fusion, the signal noise The effects of positional uncertainty are difficult to model. In order to determine the process noise coefficient of each step, in this experiment, the variance of the network output results before and after adding noise for many times is mainly used as the noise coefficient of the position. On the basis of this method, this experiment can further reflect the performance advantage of the network fusion algorithm compared to the filtering algorithm.

In order to quantitatively analyze the influence of system noise on the positioning accuracy, the simulation experiment adds different types of interference to the Bluetooth RSS input. The noise and interference in the actual scene are very complex, such as white noise, pink noise and red noise in the signal system, as well as multipath interference related to environmental factors, through-wall attenuation, etc. Therefore, in order to simplify the analysis process, this article will all interference Factors are reduced to three types:

1) Space-time constant interference. Space-time invariant noise includes many, the most typical is the dark current noise in the receiving equipment circuit. This noise is generally stable and is generally regarded as white noise. The superposition of space-time invariant noise and signal is as follows:

$$rss_w = rss + w_{tsf} \quad (17)$$

where $rss$ is the received signal strength value without interference, $w_{tsf}$ is white noise irrelevant to time and space, and the distribution of the noise obeys $N(0, \sigma_{tsf}^2)$.

2) Time-varying interference. The most typical interference over time is thermal noise, which is mainly caused by temperature changes. Temperature changes include long-term heating of internal devices and temperature changes in the external environment. To simplify this noise, this paper only considers the noise that changes periodically with the ambient temperature. The noise is superimposed as follows:

$$rss_w(t) = rss(t) + \sin(t) \cdot w_t \quad (18)$$

where $rss$ is the received signal strength value without interference, $t$ is the moment of the current received signal (in this paper, the time series data index), $w_t$ is the time-related white noise, and the noise obeys the distribution $N(0, \sigma_t^2)$.

3) Spatial interference. The interference that varies with space includes many, such as multipath interference, the interference of the fluctuation of the beacon transmitted signal itself attenuated with the distance, etc. In this paper, the interference is simplified as white noise that follows the distance. The noise is superimposed as follows:

$$rss_w = rss + \sqrt{d} \cdot w_s \quad (19)$$

where $rss$ is the received signal strength value without interference, $w_s$ is the white noise related to space, and its noise obeys the distribution $N(0, \sigma_s^2)$, $d$ is the spatial distance between the receiver and the wireless beacon.

The simulation test results of space-time invariant interference are shown in Fig. 8 and Table 2. Since the Bluetooth signal strength is generally between -30dB and -99dB, and considering that the dark current noise amplitude is small in the actual scene, the noise figure is given a small value. It can be seen from Fig. 8 that the influence of time-space invariant interference on the positioning accuracy of EKF and PF is smaller than that of the network fusion results. Since the inertial data measured by IONet is used, the complexity of inertial data noise leads to a certain gap between the results of EKF, PF and network fusion. Since only noise is added to the simulated Bluetooth signal data in the simulation results, the gap between the EKF, PF and network fusion in this group of experimental results can be regarded as the benchmark gap, so as to discuss the impact of other types of interference on the system positioning accuracy . From Fig. 8, it can also be seen that the spatial and temporal invariant interference has little effect on the positioning accuracy of EKF and PF fusion, indicating that various filtering algorithms have equivalent suppression capabilities to white noise.

Table 2 Simulation results of different noise type.

| Noise Factor | | Method | Accuracy (m) | < 80%(m) | < 95%(m) |
|---|---|---|---|---|---|
| **Space-time invariant** | $\sigma_{tsf}=1$ | EKF | 0.179 | 0.228 | 0.266 |
| | | PF | 0.216 | 0.239 | 0.580 |
| | | **Ours** | **0.078** | **0.112** | **0.160** |
| **Time variant** | $\sigma_t=1$ | EKF | 0.228 | 0.295 | 0.417 |
| | | PF | 0.216 | 0.279 | 0.460 |
| | | **Ours** | **0.117** | **0.171** | **0.249** |
| | $\sigma_t=2$ | EKF | 0.346 | 0.419 | 0.510 |
| | | PF | 0.292 | 0.351 | 0.505 |
| | | **Ours** | **0.159** | **0.232** | **0.348** |
| | $\sigma_t=4$ | EKF | 0.583 | 0.764 | 0.997 |
| | | PF | 0.499 | 0.658 | 0.914 |
| | | **Ours** | **0.248** | **0.351** | **0.630** |
| **Space variant** | $\sigma_t=1$ | EKF | 0.898 | 1.230 | 1.416 |
| | | PF | 0.861 | 1.207 | 1.449 |
| | | **Ours** | **0.118** | **0.170** | **0.255** |
| | $\sigma_t=2$ | EKF | 1.134 | 1.475 | 1.914 |
| | | PF | 0.969 | 1.293 | 1.730 |
| | | **Ours** | **0.139** | **0.210** | **0.314** |

The simulation test results of time-varying interference are shown in Fig. 9. It can be seen from Table 2 that with the increase of the noise coefficient of time-varying interference, the gap between the positioning accuracy of EKF and PF fusion and the result of network fusion gradually increases. In addition, compared with the time-space invariant interference, the positioning accuracy also decreases to a certain extent. For example, the following instability of the predicted trajectory in Fig. 9 is higher. It can also be seen from Table 2 that under the same noise figure, the positioning accuracy of EKF is 0.049 meters lower than that of space-time invariant interference.

The simulation test results of time-varying interference are shown in Fig. 10 and Table 2. It can be seen that with the increase of the noise coefficient of time-varying interference, the gap between the positioning accuracy of EKF and PF fusion and the result of network fusion gradually increases. In addition, compared with the time-space invariant interference, the positioning accuracy also decreases to a certain extent. For example, the following instability of the predicted trajectory in Fig. 10 is higher. It can also be seen from Table 2 that under the same noise figure, the positioning accuracy of EKF is 0.049 meters lower than that of space-time invariant interference. In addition, with the increase of noise coefficient, the gap between the positioning accuracy of PF and EKF fusion is also increasing, which also reflects that PF has a stronger ability to handle non-Gaussian noise.

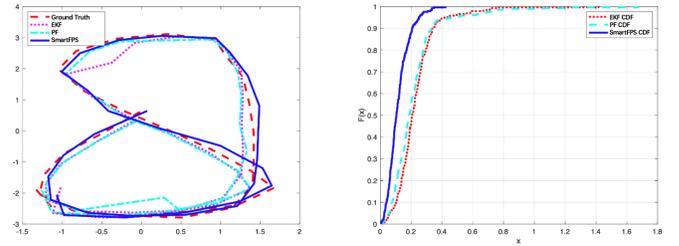

Fig. 9 Simulation results of time variant noise of $\sigma_t=1$.

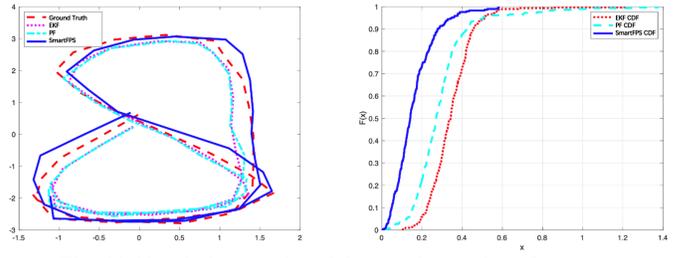

Fig. 10 Simulation results of time variant noise of $\sigma_t=2$.

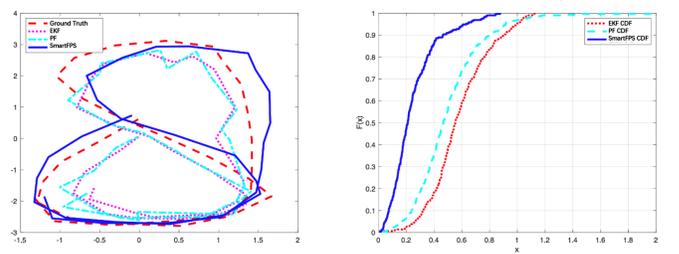

Fig. 11 Simulation results of time variant noise of $\sigma_t=4$.

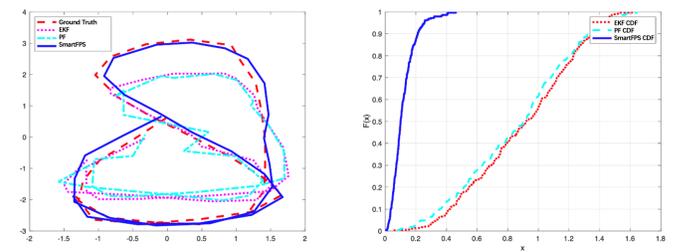

Fig. 12 Simulation results of space variant noise of $\sigma_t=1$.

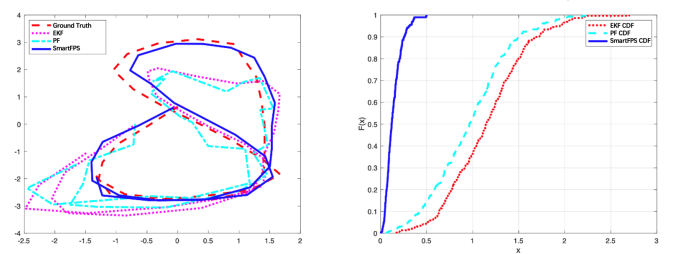

Fig. 13 Simulation results of space variant noise of $\sigma_t=2$.

### B. Experiment

The experimental verification site of this research is the middle corridor of the floor where the low-power laboratory of Southeast University, Chuangzhi Building, Pukou District, Nanjing City is located, as shown in Fig. 14. The corridor where the experimental site is located is about 30 meters long and 10 meters wide. A total of 20 Bluetooth beacons are deployed in the venue, and the distance between each Bluetooth beacon is 5-8 meters, of which No. 17-20 Bluetooth beacons are located in the low-power laboratory. The middle of the experimental site includes the stair hall and the elevator hall, so there are generally about 3-4 direct Bluetooth signals during positioning. The Bluetooth beacon adopts the E5 model of Yunli Physics. The default broadcast period of the product is 500ms. In this study, it is set to 100ms to increase the training data density. The default broadcast power of the product is 6 levels of 0dBm by default, and the coverage radius is about 50 meters. In this study, it is adjusted to 4 levels of -8dBm, and the coverage radius is 22 meters. The beacon was pasted on the wall above 1.5 meters above the ground in the experimental site.

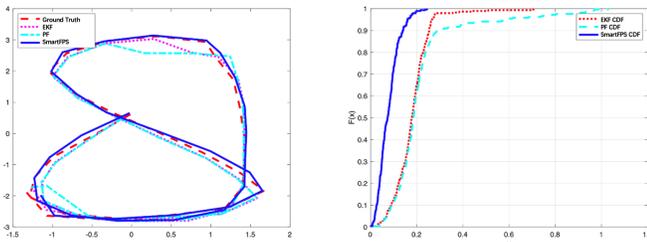

Fig. 8 Simulation results of space-time invariant noise.

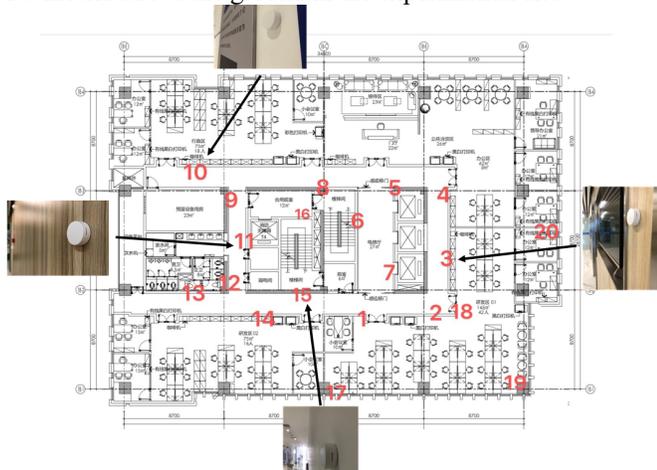

Fig. 14 Experiment Testbed

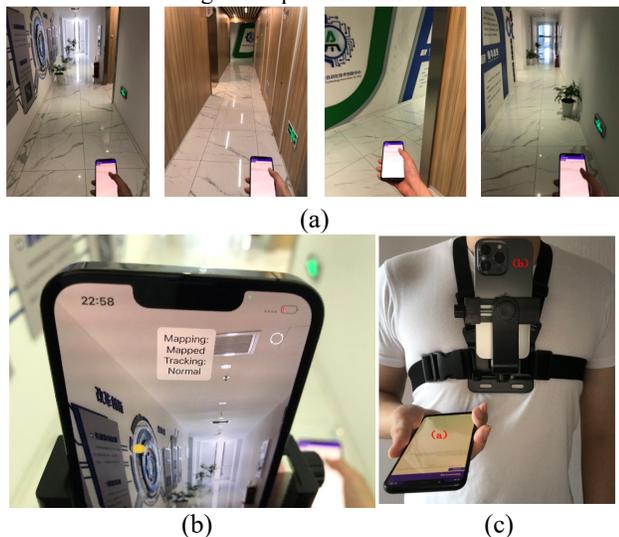

(a)

(b)                                    (c)

Fig. 15 Experiment Setup. (a) positioning scenarios of a trajectory; (b) Lidar slam on iPhone; (c) device setup on body.

In order to collect training data including accelerometer, gyroscope and Bluetooth signal strength data, Android phones including Xiaomi Mi6 and vivo x70 pro were selected as positioning mobile phones in this study. Based on the Android system, we have realized the application of real-time reading and saving of mobile phone accelerometer, gyroscope, magnetometer and other main sensor data, as well as Bluetooth signal strength data. The sampling frequency of the inertial sensor is 200Hz, and the sampling frequency of the Bluetooth signal is set to the highest in the system.

In the previous inertial navigation positioning research, the vicon system was used to accurately track the position of the mobile phone. However, considering the difficulty of deploying the system in a large scene, we imitated the research on the positioning tag acquisition system using the slam system of google tango. Different from the research, the zenfone AR Android mobile phone used in it completely relies on the combination of the rear camera to calculate the slam. This research uses the Lidar visual slam system combined with the iPhone's Lidar and the rear camera for the first time to collect the positioning label, and the positioning accuracy is higher than that of the tango system. Based on the ARkit library of iOS 15, this research implements an application that reads and saves the attitude of the mobile phone in real time. At the same time, the application will also read the real-time inertial navigation data of the iPhone to facilitate time calibration with the Android data. Fig. 15(b) shows the actual operation of the Lidar vision SLAM positioning label collection platform based on iOS ARkit. The position sampling frequency of the Lidar vision slam system is 20Hz. In order to verify the positioning accuracy of the system, we selected a rectangular line as a reference according to the texture of the floor tiles in the experimental scene, and carried out multiple continuous collections of 1-hour slam operation, walking freely throughout the whole process, and walking according to the reference line every 15 minutes. . By comparing the walking trajectories of the reference routes in different time periods, we found that the cumulative drift error of the iPhone's Lidar vision SLAM system does not exceed 20cm in 1 hour, which is a big improvement compared to the Tango system mentioned in the study, which has an error of less than 30cm in 15 minutes.

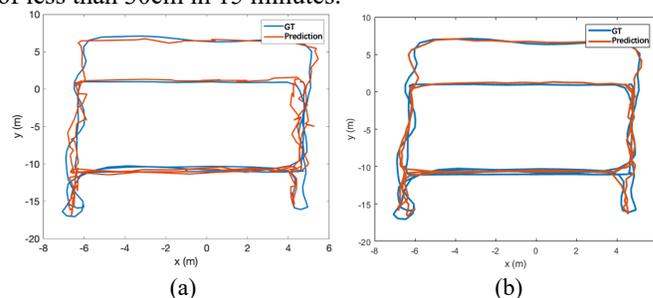

(a)                                    (b)

Fig. 16 Experiment results of a test trajectory. (a) wireless positioning encoder of SmartFPS; (b) SmartFPS.

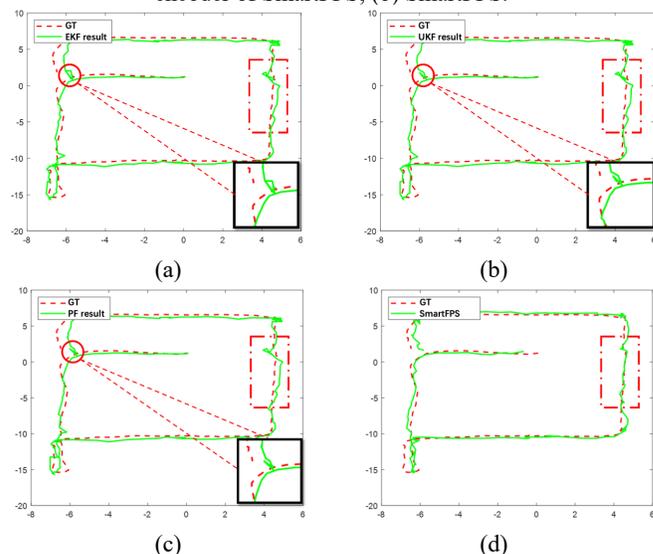

(a)                                    (b)

(c)                                    (d)

Fig. 17 Experiment results with different fusion methods of testset 1. (a) EKF; (b) UKF; (c) PF; (d) SmartFPS.

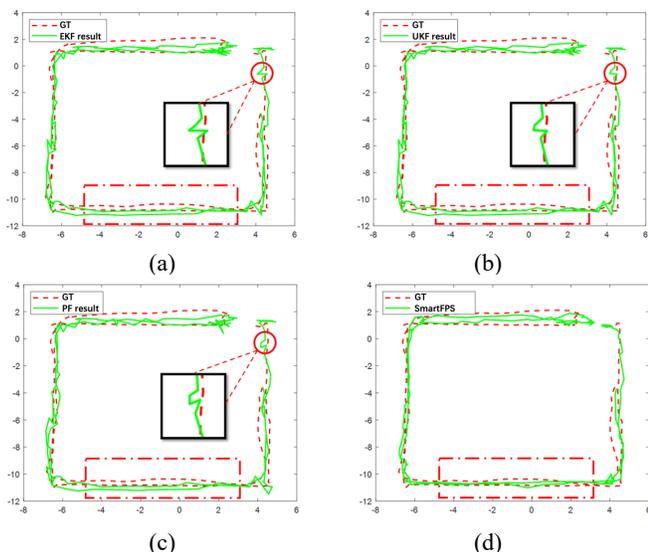

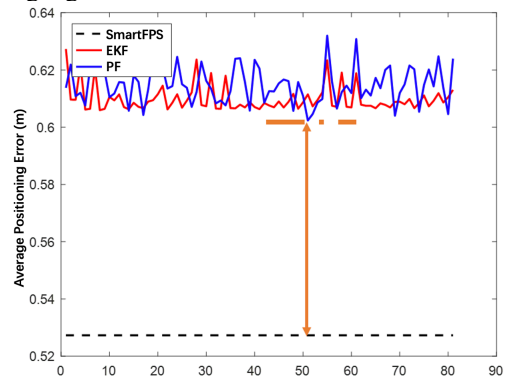

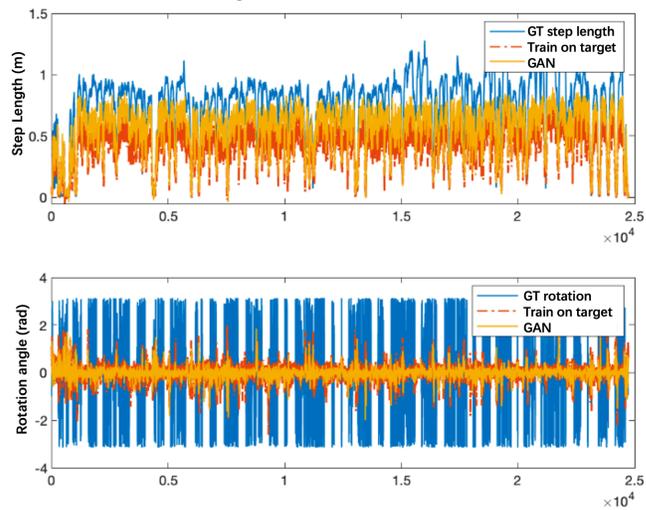

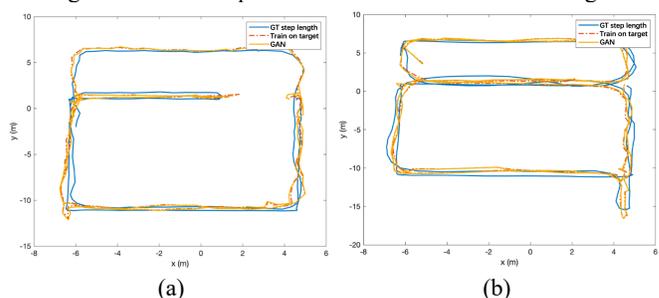

Fig. 18 Experiment results with different fusion methods of testset 2. (a) EKF; (b) UKF; (c) PF; (d) SmartFPS.

The positioning results of the implemented filtering algorithm and the network fusion algorithm proposed in this paper are shown in Fig. 17 and Fig. 18. It can be seen that the positioning trajectories of EKF and UKF are highly similar (refer to the solid line circle in the figure to mark and enlarge the part), this is because the same process noise and observation noise are used parameter. The trajectories of the results of the PF algorithm are more significantly different from those of the EKF than the UKF. The resulting trajectory of the fusion network algorithm proposed in this paper is the closest to the real trajectory (refer to the part marked by the dotted solid line in the figure). In some positions, the filtering algorithm shows large fluctuations at the same time, while the fusion network is more stable and smooth. From the table, it can be found that the performance of particle filter and unscented Kalman filter is different on different test sets. For test set 1, the location accuracy of unscented Kalman filter is slightly better than that of particle filter, while the accuracy of particle filter on test set 2 is higher. Compared with test set 1, the localization accuracy of the fusion network algorithm on test set 2 is much higher than that of the filtering algorithm. On the whole, the fusion based on neural network has obvious effect on the improvement of positioning accuracy. Compared with the best filtering algorithm, the fusion network algorithm improves the results by 13.03% and 33.80% respectively on the two test sets.

In addition, the research of this paper also finds that the process noise and observation noise coefficients calculated by the training set in the filtering algorithm are not optimal solutions, and the use of process noise and observation noise coefficients different from these statistical results has a greater impact on the final positioning accuracy. Based on the statistical results, this paper uses the global search strategy to search for the optimal solution of the process noise coefficient of the EKF and PF algorithms, and randomly adds up to 5% of the increase or decrease in the optimal solution to determine the location of the test. The accuracy results are compared with the localization results of the fusion localization network. The test results are shown in Fig. 19. It can be seen that even if the filter fusion algorithm adopts the optimal parameters, there is still a significant gap between the positioning accuracy and the fusion positioning algorithm.

Fig. 19 Filter-based fusion with global search of best variance matrix compared with SmartFPS.

Fig. 20 Domain adaptation results of inertial encoder target.

Fig. 21 Domain adaptation results of SmartFPS. (a) Testset 1; (b) Testset 2.

The test results of the inertial navigation positioning network domain adaptation algorithm are shown in Fig. 20. It can be seen that the estimated step size and deflection angle of the inertial navigation positioning network trained by the domain adaptation are closer to the real values than those estimated by the network trained directly in the source domain. The domain adaptation algorithm of test set 2 and test set 3 improves the step size estimation accuracy more significantly. The table shows that the step size estimation accuracy of test set 3 and test set 4 is improved by 63.86% and 47.74%. Overall, the domain adaptation training of inertial navigation positioning can improve the estimation accuracy of step size and deflection angle by more than 40%.

The test results of the SmartFPS domain adaptation algorithm are shown in Fig. 21. The method used here is the batch training method based on wireless coarse positioning. The update of the encoder weights of the fused localization network is performed concurrently with the update of the target domain dataset. As can be seen from Fig. 21, the localization trajectory of the fusion localization network has a certain inhibitory effect on the parameter difference, and the localization trajectory before and after domain adaptation is basically close to the real trajectory.

## VI. Conclusion

In this paper, an end-to-end neural network based wireless-inertial fusion positioning system and its transfer learning method are proposed. The experiment result shows that our method outperforms filter based methods. Our system can achieve an average positioning accuracy of 0.575 meters for different pedestrians and different mobile phones.